\documentclass[10pt,twocolumn,aps,superscriptaddress,preprintnumbers,amsmath,amssymb,nofootinbib]{revtex4}
\usepackage{epsfig}  
\usepackage{graphicx}
\usepackage{hyperref}

\hypersetup{
    colorlinks=true,
    citecolor=red,
    linkcolor=blue,
    filecolor=green,      
    urlcolor=magenta,
}
\usepackage{color}
\usepackage{float}
\usepackage{amsfonts}
\usepackage{amsmath}
\usepackage{slashed}
\usepackage{soul}
\large

\begin{document} 

\title{Experimental constraints on $\gamma$-ray strength function of 
$^{72}$Ga from $^{72}\mathrm{Ga(n,\gamma)^{72}Ga}$ capture data and reevaluation of Maxwellian-averaged cross sections of $^{71}$Ga}
\author{Sajid Ali}
\author{Rajkumar Santra}
\email{rajkumarsantra2013@gmail.com (Corresponding author)}
\author{ Gautam Gangopahyay}

\affiliation{Department of Physics, University of Calcutta, Kolkata-700009, India.}

\date{\today}


\begin{abstract}
The $\gamma$-ray strength function of medium-mass neutron rich nuclei $^{72}$Ga has been extracted from the statistical Hauser-Feshbach analysis of the available capture data of $^{71}\mathrm{Ga}(n,\gamma){}^{72}\mathrm{Ga}$ over the 0.01 - 3 MeV energy range with the required nuclear level density (NLD) of the $^{72}$Ga constraint from the work of R. Santra et al.,[\href{https://doi.org/10.1103/PhysRevC.107.064611}{Physical Review C 107, 064611 (2023)}]. The Gogny D1M model for the E1 and M1 strength functions, including low-energy upbends of $^{72}$Ga nuclei, are experimentally constrained in the present work.
Subsequently, the Maxwellian-averaged cross section(MACS) of $^{71}\mathrm{Ga}$ has been reevaluated based on the present $\gamma$-ray strength function. It is found that the present MACS value at \(kT = 30\) keV is 115.35$^{+11.92}_{-10.44}$ mb, which is consistent with previous work.

\end{abstract}

\maketitle


\section{Introduction}
The \(\gamma\)-ray strength function (\(\gamma\mathrm{SF}\)) is a statistical property of the nucleus that characterizes its electromagnetic response, and plays an important role for the statistical model descriptions of the nuclear reactions related to the photon absorption and photon emission processes in reactions \cite{Bartholomew1973,Lone1985,Goriely2019}. Detailed knowledge of \(\gamma\mathrm{SF}\) up to particle separation energy is crucial
to understand various nuclear properties, such as the existence of a Pygmy Dipole Resonance (PDR) and reactions including fusion-evaporation, spallation, and radiative capture reactions relevant to nuclear astrophysics.

There are various experimental and theoretical methods for determining the $\gamma$-ray strength function ($\gamma$SF). For example, the nuclear physics group at the Oslo Cyclotron Laboratory (OCL) introduced an experimental method \cite{schiller2000extraction} that enables the simultaneous extraction of nuclear level densities \cite{bohr1998nuclear} and $\gamma$-ray strength functions from particle–$\gamma$ coincidence measurements. Another approach through which one can constrain the $\gamma$‑ray strength function is via statistical‑model analysis of $(n,\gamma)$ or $(p,\gamma)$ capture reaction data. In the work of Netterdon \emph{et al.}, the $\gamma$‑ray strength function of $^{90}\mathrm{Zr}$ was extracted via Hauser–Feshbach \cite{PhysRev.87.366} calculations based on experimental partial cross‑section data for the $^{89}\mathrm{Y}(p,\gamma)^{90}\mathrm{Zr}$ reaction \cite{netterdon2015experimental}.In the Hauser--Feshbach (HF) model, for a typical neutron capture reaction, the required input parameters are the global optical‐model potential (OMP) for the particle channel, the nuclear level density (NLD) of ${}^{72}\mathrm{Ga}$, and the $\gamma$‑ray strength function (GSF) of compound nuclei.
. Within the Hauser–Feshbach (HF) formalism, it is assumed that the compound nucleus produced is excited to energies where its resonant states overlap extensively, and level-to-level variations become statistically averaged. \noindent
Here, \({}^{71}\mathrm{Ga}\), a mid-mass nucleus, upon capturing a neutron, forms the compound system \({}^{72}\mathrm{Ga}\), which has a high level density above its neutron separation energy \(S_n\) and possesses numerous energetically open decay channels—so that the Bohr hypothesis \cite{Bohr1936} of channel independence is satisfied for the Hauser–Feshbach statistical model \cite{Rauscher2000,PhysRev.87.366}.

For the theoretical case, the photon strength function is described using various phenomenological and microscopic frameworks. Phenomenological models are generally parameterized in terms of a Lorentzian form with giant resonance (GR) parameters. Notable examples include the Brink-Axel model \cite{BRINK1957215,PhysRev.126.671} (in which a standard Lorentzian form describes the giant dipole resonance shape) and the Kopecky-Uhl model \cite{PhysRevC.41.1941}. Microscopic PSF models, on the other hand, are built on a detailed nuclear structure theory to improve reliability and predictive capacity \cite{Wiedeking2024}. The first attempt was made by Goriely and Khan \cite{goriely2002large}, who performed large-scale quasiparticle random‑phase approximation (QRPA) calculations to obtain the $\gamma$-ray strength function.  Collective modes are described very accurately by microscopic models based on the random‑phase approximation (RPA) \cite{10.1143/PTPS.74.330}. In the quasiparticle RPA (QRPA), the pairing effect—which is important for open‑shell nuclei—is taken into account \cite{goriely2002large}. Other microscopic models include the Skyrme–Hartree–Fock–Bogoliubov model \cite{goriely2004microscopic}, the temperature-dependent relativistic mean-field model \cite{daoutidis2012large}, the Gogny–Hartree–Fock–Bogoliubov model \cite{PhysRevC.98.014327}, etc. Now the E1 and M1 $\gamma$-ray strength function calculated with mean-field-plus-QRPA still needs phenomenological corrections if we compare with experimental data \cite{goriely2002large,goriely2004microscopic,daoutidis2012large,martini2016large,goriely2016gogny}. The SF, calculated using the Gogny D1M HFB+QRPA model, reproduces the experimental SF very well, even at low energies \cite{goriely2018gogny}. In this model, the finite-range D1M Gogny force \cite{GONZALEZBOQUERA2018195} is used, and axially symmetric-deformed QRPA calculations based on Hartree-Fock-Bogoliubov (HFB) calculations are performed to obtain the strength function \cite{goriely2018gogny}.

Neutron capture on $^{71}$Ga is one of the important astrophysical reactions in the s-process, which not only determines the abundances of gallium isotopes, but also affects the abundances of the elements up to zirconium \cite{Philip1965}. Gallium is mostly produced by the weak s-process that takes place in massive stars of more than eight solar masses during convective helium-burning(at temperatures $T\sim 30$ keV) and shell carbon burning (at $T\sim 90$ keV) \cite{Philip1965}. A recent simulation indicates that a $50\%$ change in the neutron capture cross section on one gallium isotope changes the s-abundances of all the following isotopes in shell carbon burning by up to $20\%$ \cite{Anand1979}. In addition, there are three discrepant experimental results on the neutron capture cross-section on $^{71}$Ga at 25 keV \cite{Anand1979, Macklin1957, Chaubey1966} 

The main aim of the present work is twofold.
First, to investigate the $\gamma$-ray strength function
of nuclei $^{72}$Ga nuclei from statistical model analysis of $^{71}$Ga(n, $\gamma$)$^{72}$Ga capture data over the energy range 0.01 - 3 MeV. Secondly, to use the extracted $\gamma$-ray strength function of $^{72}$Ga isotope to reevaluate the Maxwell-average cross-section of $^{71}$Ga(n, $\gamma$)$^{72}$Ga capture reaction.

\section{Analysis}

\subsection{Extraction of the $\gamma$-ray strength function}
A statistical model analysis has been performed using TALYS 2.0\cite{koning2023talys} code on the available capture data\cite{tessler2022stellar, gobel2021, dovbenko1969cross, zaikin1971cross, Anand1979, PETO1967797, stavisskiį1960measurements, jiang2012measurements} of $^{72}\mathrm{Ga(n,\gamma)^{72}Ga}$ capture reaction over the energy range E$_n$ = 0.01 - 3 MeV. Here, the Hauser-Feshbach (HF) model \cite{PhysRev.87.366}, in which the required input parameters are the global optical model potential (OMP) for the $n+^{71}\mathrm{Ga}$ channel, the nuclear level density (NLD) of $^{72}\mathrm{Ga}$ and the $\gamma$ ray strength function (GSF) of $^{72}\mathrm{Ga}$. A minor variation($\leq$5\%) is observed in the capture cross-section for the different choices of the optical potential for n +$^{71}$Ga. Therefore, the global optical model potential (OMP) \cite{EfremSh_Soukhovitskii_2004} has been fixed, which is based on the Woods-Saxon form. The required NLD data of $^{72}$Ga were obtained from the experimentally measured level density \cite{PhysRevC.107.064611} of $^{72}\mathrm{Ga}$ over the excitation energy range of 3.33 to 151.33 MeV
. Since no experimental nuclear level density (NLD) data were available below 3.33 MeV, we employed the constant temperature (CT) model \cite{Gilbert1965, Brancazio1969} to estimate the level density at low excitation energies, using systematic parameters $T = 0.88$ MeV and $E_0 = -2.16$ MeV \cite{PhysRevC.72.044311}. Now the $^{71}$Ga(n, $\gamma$)$^{72}$Ga capture data are compared with the SM calculations, playing with various models for the $\gamma$-ray strength function (GSF), which are available as input options in the TALYS code. These include, for example, the Goriely hybrid model \cite{GORIELY199810}, the Kopecky-Uhl model \cite{PhysRevC.41.1941}, and the Gogny D1M HFB+QRPA model \cite{PhysRevC.98.014327}.

\begin{figure}[t]
\centering
\setlength{\unitlength}{0.05\textwidth}
\begin{picture}(10,8.5)
\put(-0.6,-1.0){\includegraphics[width=0.55\textwidth,height=0.5\textwidth, angle = 0]{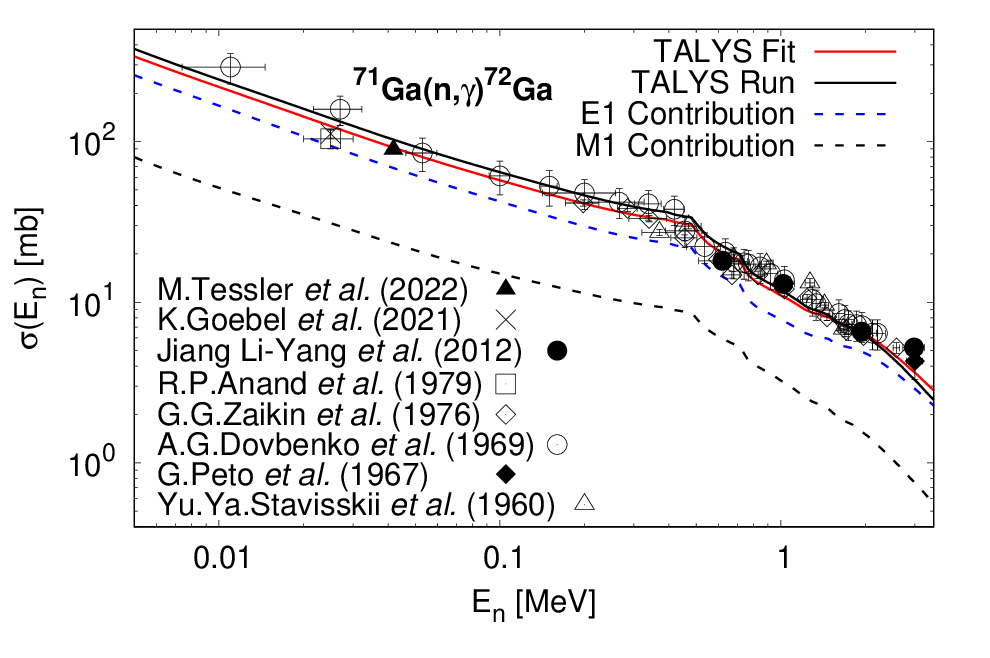}}
\end{picture}
\caption{\label{fig1} The statistical model Calculation with Gogny D1M HFB+QRPA model $\gamma$SF  for the $^{71}$Ga$(n,\gamma)$$^{72}$Ga capture cross-section and compare with available experimental data.The blue and red solid lines represent the capture cross-sections with the default Gogny model and with adjustments to $\gamma$SF parameters. The dashed blue and black lines are the E1 and M1 $\gamma$SF contribution in capture cross-sections.}
\end{figure}

A chi-square test was performed to evaluate the agreement between the simulated cross sections and the experimental data for different models of $\gamma$-SF as mentioned above \cite{tessler2022stellar,gobel2021,jiang2012measurements,anand1979kev,zaikin1971cross,dovbenko1969cross,PETO1967797,stavisskiį1960measurements}. 
Subsequent analysis indicates that the Gogny D1M HFB + QRPA model is better described (\(\chi^2\) / Ndf = 6.57) compared to other models. A comparison plot of the calculated cross section using the Gogny D1M HFB + QRPA GSF model with experimental data is shown by a solid black line in Fig.\ref{fig1}. It was found that the strength of E1 contributes 67. 63\% and the strength of M1 contributes 32.37\% in the capture cross section, as illustrated by the dashed line in Fig. \ref{fig1}.  
Furthermore, for a better description of capture data, the low-energy upbend contributions are included in the  model electric dipole (E1) and magnetic dipole (M1) strengths, including normalization in E1 part can be expressed as, 

\vspace{-0.25cm}
\small{\begin{align}
\label{eq:E1_main}
{%
 f_{E1}(\varepsilon_\gamma)
= N [ f^{\mathrm{QRPA}}_{E1}(\varepsilon_\gamma)]
\;+\;
\frac{f_0\,U}{1+\exp\!\left({\varepsilon_\gamma-\varepsilon_0}{}\right)}
},
\end{align}
\vspace{-0.5cm}
\begin{align}
\label{eq:M1_main}
{%
 f_{M1}(\varepsilon_\gamma)
= f^{\mathrm{QRPA}}_{M1}(\varepsilon_\gamma)
\;+\; C\,\exp\!\bigl(-\eta\,\varepsilon_\gamma\bigr)
},
\end{align}}

In the above expressions, $f^{\mathrm{QRPA}}_{X1}(\varepsilon_\gamma)$ denotes the dipole strength obtained from the D1M+QRPA calculation for multipolarity $X1$ ($X=E,M$) at photon energy $\varepsilon_\gamma$.  $U$ (in MeV) is the excitation energy. The symbols $N$, $f_0$, $\varepsilon_0$, $C$, and $\eta$ are free parameters. A chi-square  ($\chi^2$) analysis has been performed to minimize the total chi-square  ($\chi^2$), with using $N$, $f_0$, $\varepsilon_0$, $C$, $\eta$ as free parameters. For the global optimization of the $\chi^2$ function, the Differential Evolution (DE) method from the \texttt{scipy.optimize} module of the \texttt{SciPy} \cite{scipy} library was employed using physically motivated box bounds. Next, the best-found parameter set served as the starting point for the MIGRAD algorithm, accessed through the \texttt{iminuit} Python interface to the Minuit2 C++ library maintained by CERN \cite{m1,m2,m3}. This stage performed a gradient-based local refinement to obtain high-precision estimates and associated uncertainties. In our workflow, we set the error definition for 1$\sigma$ intervals appropriate for $\chi^2$ fits, and imposed the same box limits in Minuit as in the DE stage. The parameter uncertainties and correlation matrix were obtained from the internal covariance approximation computed by the \texttt{MIGRAD} algorithm during the minimization. Thus, this two-stage fitting procedure—employing a global search followed by local refinement—provides a robust strategy to locate the true minimum of the $\chi^2$ function. The best-fit parameters are listed in Tables \ref{tab:fit_results} and their correlation matrix shown in Fig. \ref{fig1}

\begin{table}[htbp]
\centering
\caption{Fit results from the minimization.}
\label{tab:fit_results}
\begin{tabular}{|c|c|}
\hline
Parameter & Value (fit $\pm$ 1$\sigma$) \\
\hline
$N$  & $1.29 \pm 0.02$ \\
$f_0$ (MeV$^{-4}$) & $(1.00\ \pm\ 0.13)\times10^{-10}$ \\
$\varepsilon_0$ (MeV) & $3.00 \pm 0.23$ \\
$C$ (MeV$^{-3}$) & $(1.00\ \pm\ 0.02)\times10^{-8}$ \\
$\eta$ (MeV$^{-1}$) & $0.85 \pm 0.01$ \\
\hline
$\chi^2$/dof & $5.84$ \\ 
\hline
\end{tabular}
\end{table}

\begin{figure}[t]
\centering
\setlength{\unitlength}{0.05\textwidth}
\begin{picture}(10,7.5)
\put(-0.6,-0.5){\includegraphics[width=0.55\textwidth,height=0.4\textwidth, angle = 0]{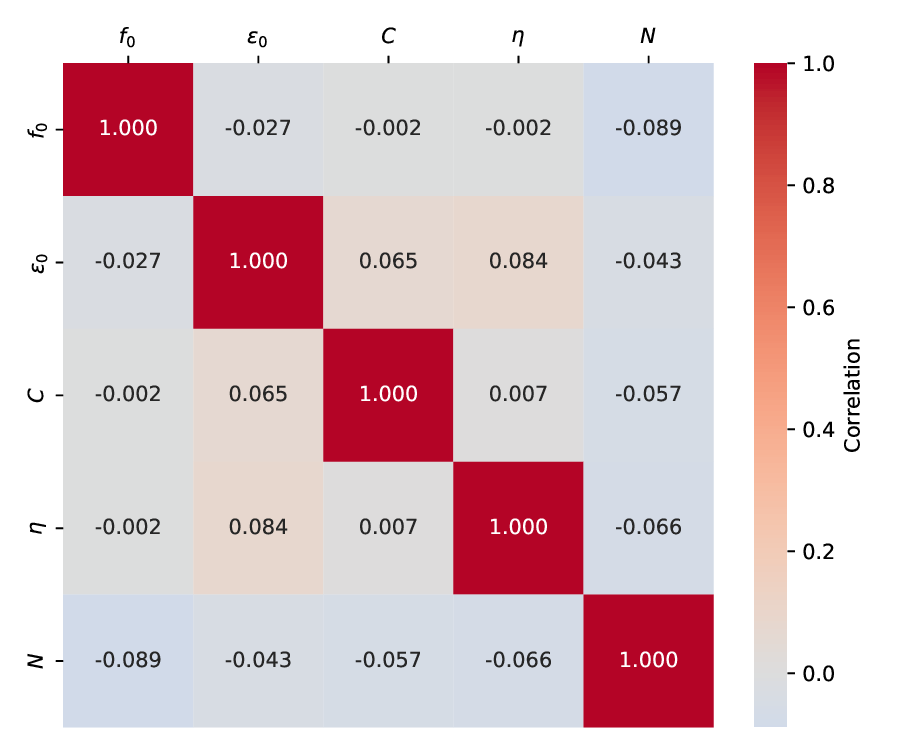}}
\end{picture}
\caption{\label{fig1} Correlation matrix of the best fitted parameters.}
\end{figure}


The resulting E1, M1, and total $\gamma$ SF values, which constitute the main outcome of this analysis, are shown in Fig. \ref{fig2}. In Fig. \ref{fig2}, the mean value of E1, M1, and the total $\gamma$SF are represented by the blue, black, and red dashed lines, respectively. In the present work, the main sources of uncertainty in $\gamma$ SF are -- first one is due to uncertainty in the $^{71}$Ga(n,$\gamma$)$^{72}$Ga capture data, and also various neutron energies (E$_n$) create an uncertainty in the low energy E1 $\gamma$SF. A typical variation of the low energy E1 $\gamma$SF with the limiting value of E$_n$ is shown in Fig.\ref{fig2}. The blue shaded band shows the resulting uncertainty band in E1 $\gamma$SF in Fig.\ref{fig2}(a) and the red band represents the uncertainty in total $\gamma$SF. Another source of uncertainty is due to the uncertainty in the input OPM \& NLD of $^{72}$Ga for the calculation of SM. The uncertainty in $\gamma$SF due to the variation of n-OPM is negligibly small, which is around 4 - 5 \%. However, due to the uncertainty of the input NLD of $^{72}$Ga, it has a dominant source of uncertainty in $\gamma$SF in the present work, which is shown in Fig.\ref{fig2}(b).
The total uncertainty in the strength function E1 is calculated considering the quadrature of these uncertainties and shown in Fig.\ref{fig2}(c). 
\subsection{Reevaluation of MACS of $^{71}$Ga}
Next,  the Maxwellian averaged cross section (MACS) for \({}^{71}\mathrm{Ga}\) is also evaluated using the resultant $\gamma$SF of the present work, and the NLD/n-OPM are the same values used for the previous part analysis. The resultant Reaction rate and MACS of $^{71}$Ga(n,$\gamma$)$^{72}$Ga capture reaction are shown in Fig. \ref{fig:figure3}(a) and (b) respectively.

\begin{figure}[t]
\centering
\setlength{\unitlength}{0.05\textwidth}
\begin{picture}(10,10)
\put(-0.2,-0.8){\includegraphics[width=0.55\textwidth,height=0.5\textwidth, angle = 0]{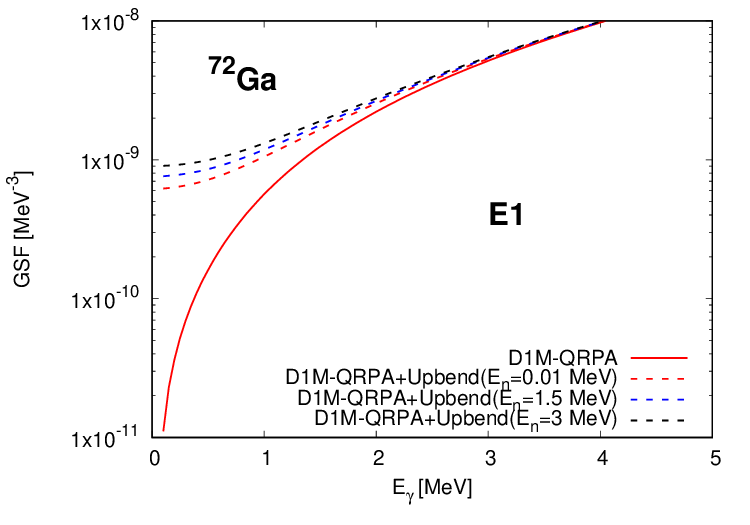}}
\end{picture}
\caption{\label{fig2} The variations of upbend contribution
of E1 $\gamma$SF with neutron energy.  }
\end{figure}

\begin{figure}[t]
\centering
\setlength{\unitlength}{0.05\textwidth}
\begin{picture}(10,15)
\put(-0.2,-0.5){\includegraphics[width=0.5\textwidth,height=0.8\textwidth, angle = 0]{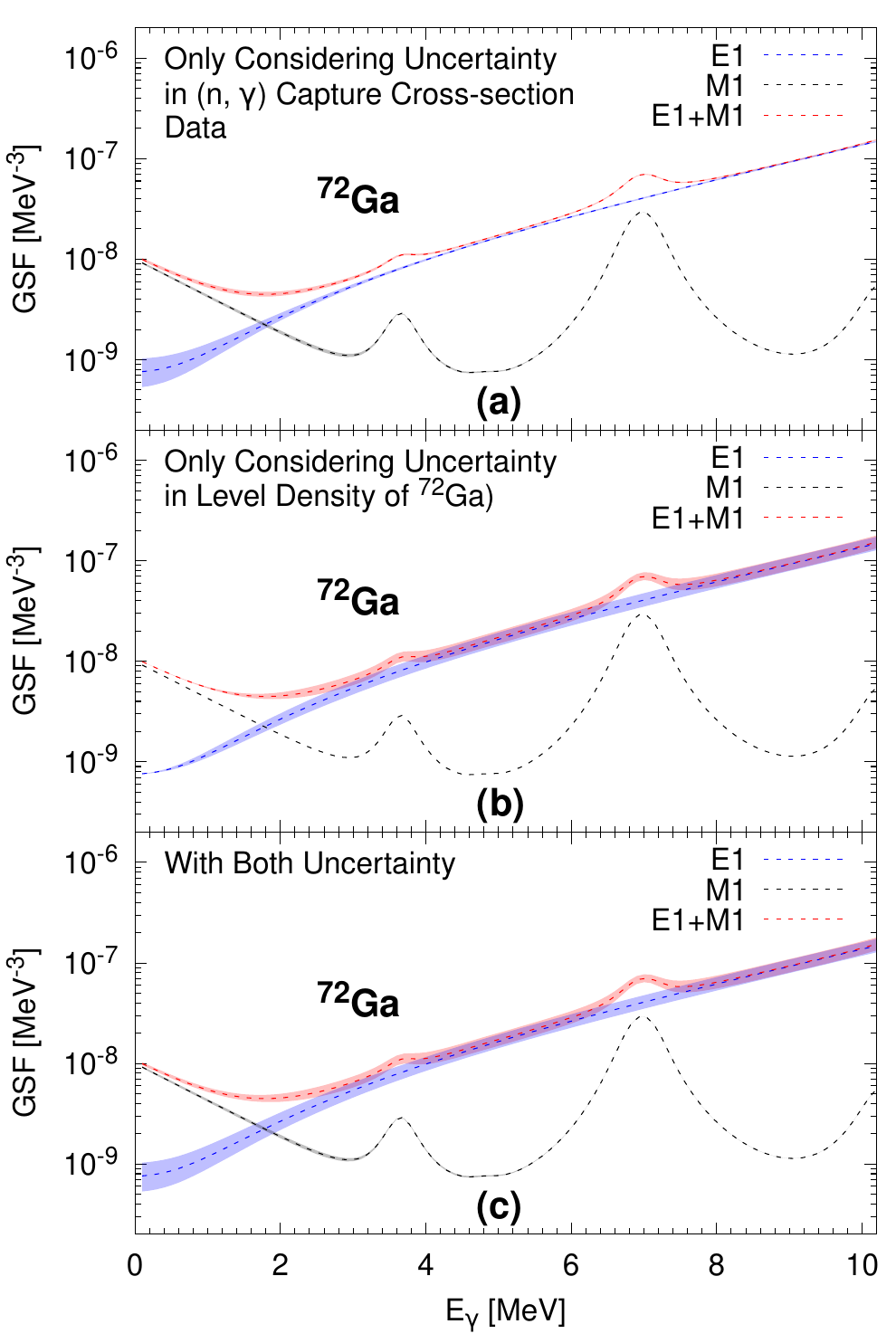}}
\end{picture}
\caption{\label{fig2} (a) The extracted $\gamma$SF with considering only (n, $\gamma$) capture data uncertainty.(b) The extracted $\gamma$SF with considering only the uncertainty in NLD of $^{72}$Ga. (c) The $\gamma$SF considering both uncertainties.}
\end{figure}

\begin{table}[htbp]
\centering
\caption{Comparison of Maxwellian Averaged Cross Section (MACS) values for $^{71}$Ga at $kT = 30$ keV from different references.}
\label{tab:MACS_71Ga}
\begin{tabular}{clc}
\hline
\hline
\textbf{No.} & \textbf{Reference} & \textbf{MACS (mb)} \\
\hline
1 & KADoNiS v3.0 (2009) \cite{macs1}        & 123(8) \\[3pt]
2 & Anand \textit{et al.} (1979) \cite{macs2}                      & 79(23) \\[3pt]
3 & Walter \textit{et al.} (1986) \cite{macs3}                      & 130(8) \\[3pt]
4 & Tessler \textit{et al.} (2022) \cite{tessler2022stellar} &  105(7) \\[3pt]
5 & Allen \textit{et al.} (1971) \cite{macs5} & 120(30) \\[3pt]
6 & Chadwick \textit{et al.} (2006) \cite{macs6} & 122.3 \\[3pt]
7 & Shibata \textit{et al.} (2002) \cite{macs7} & 103.2 \\[3pt]
8 & Rauscher \textit{et al.} (2000) \cite{macs8} & 86 \\[3pt]
9 & Harris \textit{et al.} (1981) \cite{macs9} & 93 \\[3pt]
10 & Woosley \textit{et al.} (1978) \cite{macs10} & 50 \\[3pt]
11 & Goriely \textit{et al.} (2002) \cite{macs11} & 131 \\[3pt]
12 & Goriely \textit{et al.} (2005) \cite{macs12} & 117 \\[3pt]
\hline
\hline
\end{tabular}
\end{table}

\begin{figure}[t]
\centering
\setlength{\unitlength}{0.05\textwidth}
\begin{picture}(10,8)
\put(-0.2,-.5){\includegraphics[width=0.52\textwidth,height=0.45\textwidth, angle = 0]{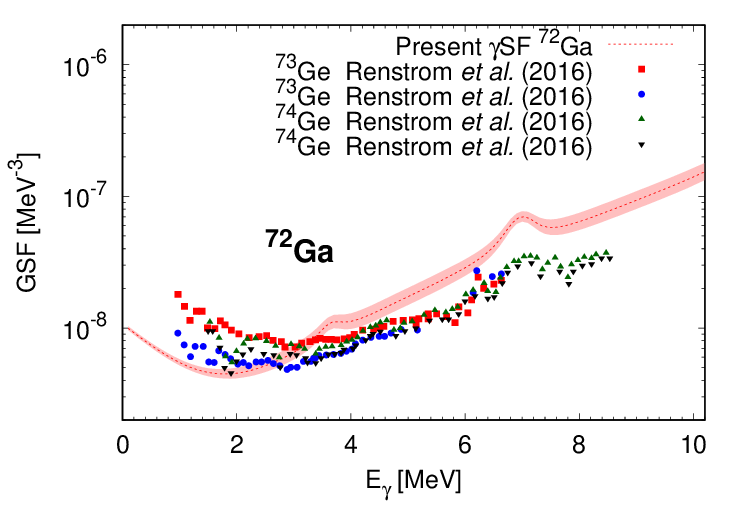}}
\end{picture}
\caption{\label{fig3} The present $\gamma$SF of $^{72}$Ga compare with $^{73,74}$Ge \cite{ren2016} $\gamma$SF which available in literature. }
\end{figure}


\begin{figure}[t]
\centering
\setlength{\unitlength}{0.05\textwidth}
\begin{picture}(10,8)
\put(-0.2,-0.6){\includegraphics[width=0.5\textwidth,height=0.38\textwidth, angle = 0]{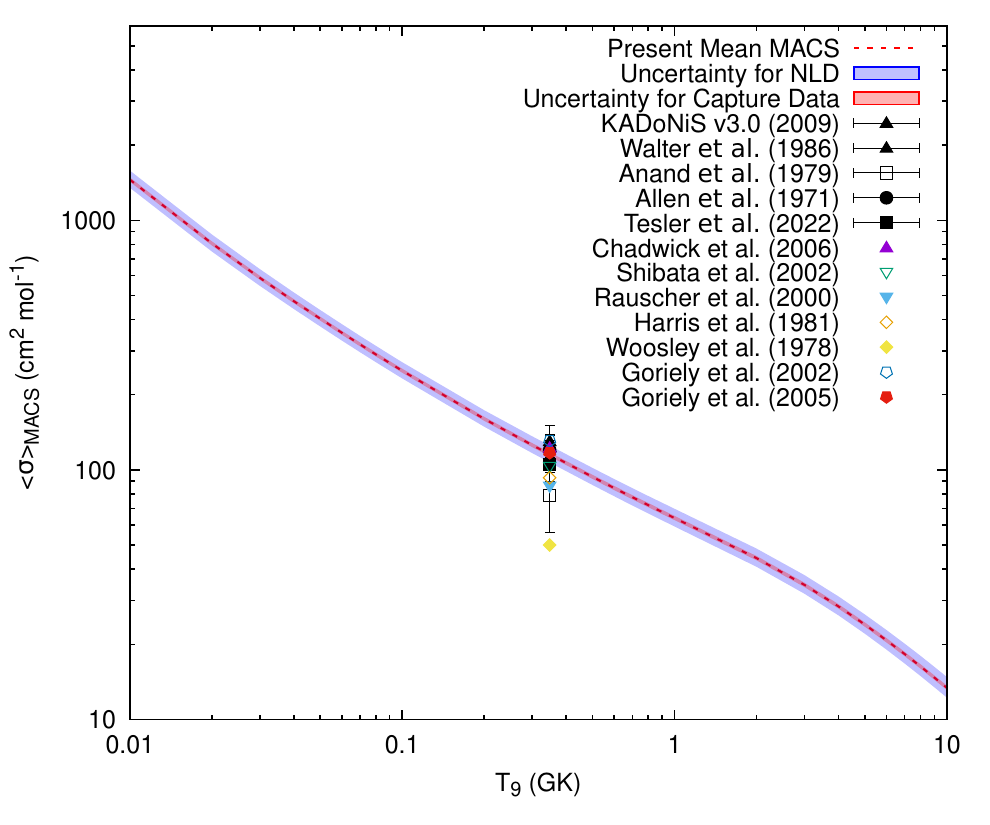}}
\end{picture}
\caption{\label{fig:figure3} The reaction rate and MACS cross-section variation with the steller temperature of the $^{71}$Ga$(n,\gamma)$$^{72}$Ga capture reaction}
\end{figure}

\section{Results and Discussion}
In our analysis, we found that the Gogny D1M HFB + QRPA model of the $\gamma$-ray strength function provided the best description of the experimental neutron capture cross sections compared to other $\gamma$-SF models. The available capture of data for the reaction $^{71}$Ga(n,$\gamma$)$^{72}$Ga is well described with SM calculations by adjusting the magnitude of E1 $\gamma$SF of the Gogny D1M HFB + QRPA model and the parameters of the low-energy up-bend contributions in $\gamma$SF. The Gogny D1M HFB + QRPA strength function model of $^{72}$Ga has been experimentally constrained from the SM descriptions of available $^{71}$Ga(n,$\gamma$)$^{72}$Ga capture data. The uncertainty in $\gamma$SF arises primarily from the error in the input NLDs of $^{72}$Ga, which is significantly greater than the uncertainty in $\gamma$SF caused by the capture data. The $\gamma$SF of $^{72}$Ga in the present work is compared with the experimental data available of $^{73,74}$Ge \cite{ren2016} as shown in Fig. \ref{fig3}(b). The present value of $\gamma$SF matches in the energy range of 2 - 3 MeV with the data of $^{73,74}$Ge. However, the overall slope of the present $\gamma$SF of $^{72}$Ga is slightly higher than that of the $^{73,74}$Ge data.\\
 Using our resultant  $\gamma$SF and the experimental value of the NLD of $^{72}$Ga of R. Santra et al.[\href{https://doi.org/10.1103/PhysRevC.107.064611}{Physical Review C 107, 064611 (2023)}], we have also re-evaluated the reaction rate and the Maxwellian-averaged cross section (MACS) for \({}^{71}\mathrm{Ga}\), as shown in Fig. \ref{fig:figure3}, upper and lower, respectively. The present MACS value at \(kT = 30\) keV is compared with previous work reported in literature\cite{} as shown in the lower panel of Fig. \ref{fig:figure3} and Table \ref{tab:MACS_71Ga}. It was found that the MACS values at \(kT = 30\) keV, as reported in previous works, have a wide range, as listed in Table \ref{tab:MACS_71Ga}. However, in the present study, the MACS at an energy of \(kT = 30\) keV yields a value of 115.35$^{+11.92}_{-10.44}$ mb, which is consistent with most of the previous works\cite{tessler2022stellar,macs1,macs2,macs3,macs5,macs6,macs7,macs8, macs9,macs10,macs11,macs12}.     
 
\section{Acknowledgments}
S.A. gratefully acknowledges the Research Fellowship from the Human Resource Development Group, Council of Scientific \& Industrial Research (CSIR–HRDG), India, vide reference no. 09/0028(16241)/2023‑EMR‑I. Rajkumar Santra is grateful to the Science and Engineering Research Board/Anusandhan National Research Foundation (SERB/ANRF, India) for financial support under the National Postdoctoral Fellowship (N-PDF) scheme, vide reference no. PDF/2023/001152.


\begin{thebibliography}{100}

\bibitem{Bartholomew1973} G.A. Bartholomew et al, Adv. Nucl. Phys., 7:229, 1973.

\bibitem{Lone1985}
M. A. Lone.
    page 238. D. Reidel, 1986.

\bibitem{Goriely2019}
St{\'e}phane Goriely et al.
         The European Physical Journal A, 55:1--52, 2019.

\bibitem{schiller2000extraction}
A Schiller et al.
         Physics Research Section A, 447(3):498--511, 2000.

\bibitem{bohr1998nuclear}
Aage Niels Bohr and Ben R Mottelson.
         Nuclear Structure.
    1998.

\bibitem{PhysRev.87.366}
Walter Hauser and Herman Feshbach.
         Phys. Rev., 87:366--373, Jul 1952.

\bibitem{netterdon2015experimental}
Lars Netterdon et al.
         Physics Letters B, 744:358--362, 2015.

\bibitem{Bohr1936}
N. Bohr.
         Nature, 137:344--346, 1936.

\bibitem{Rauscher2000}
T. Rauscher and F.-K. Thielemann.
         At. Data Nucl. Data Tables, 75:1--351, 2000.

\bibitem{BRINK1957215}
D.M. Brink.
         Nuclear Physics, 4:215--220, 1957.

\bibitem{PhysRev.126.671}
Peter Axel.
         Phys. Rev., 126:671--683, Apr 1962.

\bibitem{PhysRevC.41.1941}
J. Kopecky and M. Uhl.
         Phys. Rev. C, 41:1941--1955, May 1990.

\bibitem{Wiedeking2024}
M. Wiedeking and S. Goriely.
         Royal Society A, 382(2275):20230125, Jul 2024.

\bibitem{goriely2002large}
St{\'e}phane Goriely and Elias Khan.
         Nuclear Physics A, 706(1-2):217--232, 2002.

\bibitem{10.1143/PTPS.74.330}
Nguyen Van Giai.
         Progress of Theoretical Physics Supplement, 74-75:330--341, 01 1983.

\bibitem{goriely2004microscopic}
St{\'e}phane Goriely et al.
         Nuclear Physics A, 739(3-4):331--352, 2004.

\bibitem{daoutidis2012large}
Ioannis Daoutidis and St{\'e}phane Goriely.
         Physical Review C, 86(3):034328, 2012.

\bibitem{PhysRevC.98.014327}
S. Goriely et al.
         Phys. Rev. C, 98:014327, Jul 2018.

\bibitem{martini2016large}
Marco Martini et al.
         Physical Review C, 94(1):014304, 2016.

\bibitem{goriely2016gogny}
St{\'e}phane Goriely et al.
         Physical Review C, 94(4):044306, 2016.

\bibitem{goriely2018gogny}
St{\'e}phane Goriely et al.
         Physical Review C, 98(1):014327, 2018.

\bibitem{GONZALEZBOQUERA2018195}
C. Gonzalez-Boquera et al.
         Physics Letters B, 779:195--200, 2018.

\bibitem{Philip1965}
Philip A. Seeger and William A. Fowler.
         Astrophysical Journal Supplement, 11(1):121, 1965.

\bibitem{Anand1979}
D. Bhattacharya R. Anand, M. Jhingan and E. Kondaiah.
         Nuovo Cimento, 50A:247, 1979.

\bibitem{Macklin1957}
N. H. Lazar R. L. Macklin and W. S. LYON.
         Phys. Rev.,, 107:504, 1957.

\bibitem{Chaubey1966}
A. K. Chaubey and M. L. SehgaL.
         Phys. Rev.,, 152:055, 1966.

\bibitem{koning2023talys}
Arjan Koning et al.
         The European Physical Journal A, 59(6):131, 2023.

\bibitem{tessler2022stellar}
M Tessler et al.
         Physical Review C, 105(3):035801, 2022.

\bibitem{gobel2021}
Kathrin G{\"o}bel et al.
         Phys. Rev. C, 103:025802, Feb 2021.

\bibitem{dovbenko1969cross}
AG Dovbenko et al.
         Soviet Atomic Energy, 26(1):82--85, 1969.

\bibitem{zaikin1971cross}
GG Zaikin et al.
         Ukrainian Physics Journal, 16:1476, 1971.

\bibitem{PETO1967797}
G. Petö et al.
         Journal of Nuclear Energy, 21(10):797--801, 1967.

\bibitem{stavisskiį1960measurements}
Yu Stvi et al, Journal of Nuclear Energy, 12(10):`132--133, 1960.

\bibitem{jiang2012measurements}
Liyang Jiang et al.
         Atomic Energy Science and Technology, 46(6):641--647, 2012.

\bibitem{EfremSh_Soukhovitskii_2004}
Efrem Sh Soukhovitskii et al.
         Journal of Physics G, 30(7):905, may 2004.

\bibitem{PhysRevC.107.064611}
Rajkumar Santra et al.
         Phys. Rev. C, 107:064611, Jun 2023.

\bibitem{Gilbert1965}
A. Gilbert and A. G. W. Cameron.
         Canadian Journal of Physics, 43:1446, 1965.

\bibitem{Brancazio1969}
P. J. Brancazio and A. G. W. Cameron.
         Canadian Journal of Physics, 47:1029, 1969.

\bibitem{PhysRevC.72.044311}
Till von Egidy and Dorel Bucurescu.
         Phys. Rev. C, 72:044311, Oct 2005.

\bibitem{GORIELY199810}
S. Goriely.
         Physics Letters B, 436(1):10--18, 1998.

\bibitem{anand1979kev}
RP Anand et al.
         Nuovo Cimento. A, 50(2):247--257, 1979.

\bibitem{scipy}
     Nature methods, 17(3):261--272, 2020.

\bibitem{m1}
F. James and M. Roos.
         Computer Physics Communications, 10(6):343--367, 1975.

\bibitem{m2}
Fred James.
    Cern, 1994.

\bibitem{m3}
Hans Dembinski et al.
    "scikit-hep/iminuit", oct 2025.

\bibitem{ren2016}
T. Renstrøm et. al;Phys. Rev. C, 93:064302, June 2016.

\bibitem{macs1} I. Dillmann et al.
    JRC-IRMM, Geel, Belgium, 2009.
    \url{http://www.kadonis.org/}.

\bibitem{macs2}
RP Anand et al.
         Il Nuovo Cimento A (1971-1996), 50(2):247--257, 1979.

\bibitem{macs3}
G Walter, H Beer, F K{\"a}ppeler, G Reffo, and F Fabbri.
    The s-process branching at se-79.
         Astronomy and Astrophysics (ISSN 0004-6361), vol. 167, no. 1, Oct. 1986, p. 186-199., 167:186--199, 1986.

\bibitem{macs5}
BJ Allen et al.
    In      Advances in Nuclear Physics: Volume 4, pages 205--259. Springer, 1971.

\bibitem{macs6}
MB Chadwick et al. Nuclear data sheets, 107(12):2931--3060, 2006.

\bibitem{macs7}
Keiichi Shibata et al.
         Journal of nuclear science and technology, 39(11):1125--1136, 2002.

\bibitem{macs8}
Thomas Rauscher et al.
         arXiv preprint astro-ph/0004059, 2000.

\bibitem{macs9}
MJ Harris Astrophysics and Space Science, 77(2):357--367, 1981.

\bibitem{macs10}
SE Woosley et al. Atomic Data and Nuclear Data Tables, 22(5):371--441, 1978.

\bibitem{macs11}
S Goriely, 2002.
    Hauser-Feshbach rates for neutron capture reactions (version 9/12/2002).

\bibitem{macs12}
S Goriely. Technical report, 2005.
    Hauser-Feshbach rates for neutron capture reactions (version 8/29/2005).

\end{thebibliography}
\end{document}